\documentclass[%
 aip,
 amsmath,amssymb,
 reprint,%
]{revtex4-1}

\usepackage{graphicx}
\usepackage{dcolumn}
\usepackage{bm}

\usepackage[utf8]{inputenc}
\usepackage[T1]{fontenc}
\usepackage{mathptmx}
\usepackage{etoolbox}
\usepackage{xcolor}

\newcommand{\be}{\begin{equation}}
\newcommand{\ee}{\end{equation}}
\newcommand{\beqn}{\begin{eqnarray}}
\newcommand{\eeqn}{\end{eqnarray}}

\makeatletter
\def\@email#1#2{%
 \endgroup
 \patchcmd{\titleblock@produce}
  {\frontmatter@RRAPformat}
  {\frontmatter@RRAPformat{\produce@RRAP{*#1\href{mailto:#2}{#2}}}\frontmatter@RRAPformat}
  {}{}
}%
\makeatother
\begin{document}

\preprint{AIP/123-QED}

\title{Finite-size scaling and dynamics in a two-dimensional lattice of identical oscillators with frustrated couplings}
\author{Róbert Juhász}
\email{juhasz.robert@wigner.hun-ren.hu}
 \affiliation{HUN-REN Wigner Research Centre for Physics, Institute for Solid State Physics and Optics, H-1525 Budapest, P.O.Box 49, Hungary}

\author{Géza Ódor}
\affiliation{HUN-REN Centre for Energy Research, Institute of Technical Physics and Materials Science, H-1525 Budapest, P.O.Box 49, Hungary}

\date{\today}

\begin{abstract}
A two-dimensional lattice of oscillators with identical (zero) intrinsic frequencies and Kuramoto type of interactions with randomly frustrated couplings is considered. Starting the time evolution from slightly perturbed synchronized states, we study numerically the relaxation properties, as well as properties at the stable fixed point which can also be viewed as a metastable state of the closely related XY spin glass model.  
According to our results, the order parameter at the stable fixed point shows generally a slow, reciprocal logarithmic convergence to its limiting value with the system size. The infinite-size limit is found to be close to zero for zero-centered Gaussian couplings, whereas, for a binary $\pm 1$ distribution with a sufficiently high concentration of positive couplings, it is significantly above zero. Besides, the relaxation time is found to grow algebraically with the system size. Thus, the order parameter in an infinite system approaches its limiting value inversely proportionally to $\ln t$ at late times $t$, similarly to that found in the model with all-to-all couplings [Daido, Chaos {\bf 28}, 045102 (2018)]. As opposed to the order parameter, the energy of the corresponding XY model is found to converge algebraically to its infinite-size limit.
\end{abstract}

\maketitle


\begin{quotation}
We consider a dynamical system composed of oscillators (or rotators) arranged in a two dimensional lattice with random, nearest-neighbor couplings of both positive and negative signs. The time derivative of phases is governed by the negative gradient of a potential function, which is the energy of the classical XY spin glass model. We aimed to investigate how the frustration present in the system due to attractive and repulsive couplings affects dynamics. Starting the system from almost ordered (synchronized) states, the time evolution and fixed-point value of the order parameter and other quantities were studied numerically.    
We find that the order parameter at the stable fixed point, which is a metastable state of the corresponding XY model, converges very slowly, proportionally to $1/\ln L$ to its infinite-size limit. The latter can be close to or significantly higher than zero,  depending on the distribution of couplings. The relaxation time is found to grow algebraically 
with the system size. These results imply that the superslow, $O(1/\ln t)$ relaxation of the order parameter observed earlier in the model with all-to-all couplings (Daido, 2018), appears also in dimensions as low as $d=2$. 
\end{quotation}

\section{\label{sec:intro}Introduction}

Synchronization of interacting degrees of freedom is a ubiquitous phenomenon in a broad palette of natural and artificial systems \cite{pikovsky,acebron,ARENAS200893} such as the human brain \cite{Deco3366}, power grids~\cite{ARENAS200893}, or chemical reactions~\cite{Kissrev}, to mention a few. 
A paradigmatic theoretical model of these phenomena is the Kuramoto model \cite{kuramoto,kura}, which describes a set of interacting oscillators, the phases $\varphi_i$ of which evolve according to the equations 
\be 
\dot{\varphi}_i(t) = \omega_i + \sum_{j}K_{ij}\sin(\varphi_j-\varphi_i), \quad i=1,2,\dots,N.
\label{gen}
\ee
Here, the intrinsic frequencies $\omega_i$ are independent, identically distributed (i.i.d.) quenched random variables. 
In the simplest non-trivial case, where the couplings are positive and  uniform ($K_{ij}=K>0$), the interaction tends to synchronize the phases of coupled oscillators,  concurring thereby with the effect of intrinsic frequencies. This leads to the existence of a synchronization transition at a critical coupling strength $K_c$ at and above the lower critical dimension $d_l=4$ \cite{chate_prl,ssk,HPClett}, including the analytically tractable all-to-all coupled case where $K_c\sim N^{-1}$, $N$ denoting the number of oscillators \cite{kuramoto,ott}.

In the general model in Eq. (\ref{gen}), an other type of frustration can also be realized by allowing for the existence of random couplings of both positive and negative signs. Negative couplings enforce a phase difference of $\pi$ rather than $0$ to coupled oscillators, hence introducing a frustration, even in the absence of intrinsic frequencies ($\omega_i=0$). Such a generalization of the model may be motivated by the well-known existence of inhibitory connections in the  brain\cite{Wilson1972ExcitatoryAI,10.1162/089976603321043685,PhysRevLett.120.244101,Flycikk}.
The late-time behavior of this model with frustrated, symmetric ($K_{ij}=K_{ji})$ all-to-all couplings has recently been studied for continuous \cite{daido}, as well as discrete \cite{hong_martens} distributions of couplings. For a zero-centered normal distribution, the time-dependence of the Kuramoto order parameter was numerically studied starting from slightly perturbed synchronized states, and a superslow relaxation of the form $R(t)\sim 1/\ln(t/t_0)$ was observed \cite{daido}. In the same model but with a bimodal ($K_{ij}=\pm 1$) distribution of couplings, the $t\to\infty$ limiting value of the order parameter (starting the system from completely unsynchronized states) was found to show a step-like discontinuity as a function of the concentration $c$ of negative bonds, at $c_c=0.5$ in the large system size limit \cite{hong_martens}.    
In this work, we aim at extending these studies on the relaxation in the  presence of random frustrated couplings to two dimensions, which is a the lowest dimension where frustration appears.   

The model in Eq. (\ref{gen}) in the absence of intrinsic frequencies ($\omega_i=0$) and with symmetric couplings ($K_{ij}=K_{ji}$), such as the frustrated models of Refs. \cite{daido} and \cite{hong_martens} is well-known to have a close relationship with the classical XY spin model, as it can be recast in the form 
\be 
\dot\varphi_i(t) = -\frac{\partial U}{\partial\varphi_i},
\quad i=1,2,\dots,N,
\label{grad}
\ee
where 
\be
U(\{\varphi_i\})=-\sum_{i<j}K_{ij}\cos(\varphi_i-\varphi_j)
\label{UXY}
\ee
is the energy of the XY spin model, and the phases $\varphi_i$ correspond to the angles of spins.
Due to the form of Eq. (\ref{grad}), the energy $U(\{\varphi_i(t)\})$ is a non-increasing function of $t$ during the time evolution, and in the limit $t\to\infty$, the trajectory runs to a stable fixed-point ($\dot\varphi_i=0$) which is a local minimum of the energy surface of the XY model.   
We note that the trajectory $\{\varphi_i(t)\}$ obeying the evolution equation in Eq. (\ref{grad}) for some initial condition $\{\varphi_i(0)\}$ can also be regarded as a continuous-time limit path of the gradient descent, a method of minimizing differentiable, multivariate functions \cite{cauchy}.

Concerning the subject of this work, the two-dimensional frustrated Kuramoto  model, the evolution equations (\ref{grad}) drive the system, in general, to a metastable state of the corresponding two-dimensional XY spin glass model. The ground-state properties of this model have been addressed in several works and, although these may be different from properties of metastable states considered in this paper, they serve as a reference point to our study. Finding the ground state of XY spin glasses is computationally a non-polynomial problem, hence studies resorted to approximations \cite{vannimenus} or numerical techniques limited to relatively small linear system sizes $L$~\cite{gawiec,weigel}. For this reason, the question whether the ground state has a non-zero magnetization (which is equivalent to the Kuramoto order parameter) in the $L\to\infty$ limit has not settled yet. For binary ($\pm 1$) disorder, the coherent potential approximation \cite{vannimenus} predicted the ferromagnetic (or synchronized) state to be unstable for any $c>0$ concentration of negative bonds, although the interpretation of this result as a transition from a ferromagnetic to a paramagnetic state or rather a crossover is unclear.  
For the same model, by an improved algorithm \cite{gawiec} based on the successive reorientation of spins to the direction of local fields \cite{walker}, sets of low-energy quasi-degenerate states were generated, and among others, magnetic properties were analyzed for linear system sizes $L\le 80$. The data were consistent with a spin-glass state with zero magnetization for any $c>0$, but, due to the poor convergence with increasing system size, ordering could not be ruled out for sufficiently small values of $c$ \cite{gawiec}. A recent work used the genetic embedding matching technique to find the true ground state with a high reliability for small systems of size $L\le 28$ with binary disorder \cite{weigel}. The ground state was found to be unique (up to a global $O(2)$ rotation symmetry) and the morphology of almost degenerate excited states was revealed, but, owing to the limited system sizes, the magnetization in the $L\to\infty$ limit remained elusive also in this work.  

As aforementioned, rather than finding the ground state, we consider in this paper metastable states, as well as the relaxation toward them, which are accessible by the evolution equations (\ref{grad}), starting from slightly perturbed synchronized initial states. 
Since the starting configuration is magnetic, having an $R\approx 1$ order parameter, one may expect that, at the fixed point, some non-zero remanent magnetization freezes in the system, as a memory of the initial magnetic state. 
According to our numerical results obtained for system sizes as large as $L=4096$, this seems indeed to be the case, at least for binary disorder with a moderate concentration of negative bonds. As opposed to this, for zero-mean, normally distributed couplings, the order parameter seems to vanish at the fixed point in the limit $L\to\infty$. 
In both cases, convergence is rather slow with increasing system size, fitting well to a logarithmic dependence of the form $R(L)-R(\infty)\sim 1/\ln L$. 
On the other hand, the energy density of the corresponding XY model, as well as the relaxation time exhibit an algebraic dependence on the system size. 
The time-dependent order parameter in an infinite system, as it also follows from the above results, decays to its limiting value proportionally to $1/\ln t$, similar to that observed in Ref. \cite{daido} in the all-to-all coupled model with normally distributed couplings. 

A rest of the paper is organized as follows. The model and the method are described in details in Sec. \ref{sec:model}. The numerical results on the fixed-point properties, as well as on the dynamics are presented in Sec. \ref{sec:results}. Finally, the results are discussed in Sec. \ref{sec:discussion}.

\section{The model and methods}
\label{sec:model}

In this paper, we consider the Kuramoto model without intrinsic frequencies on two-dimensional lattices of size $L\times L$, with nearest-neighbor interactions and periodic boundary condition in both directions. The evolution equations of this model can be formulated in terms of the energy function $U(\{\varphi_\})$ as given in Eqs. (\ref{grad}) and (\ref{UXY}). The couplings $K_{ij}$ are symmetric, i.i.d. quenched random variables taken either from a zero-mean and unit-variance Gaussian distribution or from a bimodal distribution
\be 
P(K_{ij})=(1-c)\delta(K_{ij}-1) + c\delta(K_{ij}+1),
\ee
where $0\le c\le 1$ denotes the concentration of negative bonds. 
The system is started from random initial states in which the phases are i.i.d. random variables drawn from a uniform distribution on the domain $[-\phi_0/2,\phi_0/2]$, with $\phi_0=0.1$ if not specified otherwise. Note that, due to $\sum_i\dot\varphi_i(t)=0$, the sum of phases $\sum_i\varphi_i(t)$ is a constant of motion, which, due to the global $O(2)$ symmetry of the model, can be set to $0$ without the restriction of generality. 
Then the time-evolution of the system was followed by numerically solving Eqs. (\ref{grad}) by the fourth-order Runge-Kutta method \cite{NumR} with a fixed step length of $0.05$. For system sizes $L > 512$, we applied a GPU implementation of the Bulirsch-Stoer algorithm \cite{NumR}. 
During the time evolution, we measured the Kuramoto order parameter 
\be 
R(t)=\frac{1}{L^2}\left|\sum_{j=1}^{L^2}e^{i\varphi_j(t)}\right|,
\label{Rt}
\ee
which is equivalent to the magnetization of the XY model
and, the underlying lattice being bipartite (for even $L$), we also monitored the sublattice order parameter, $R_s$, in which the summation of Eq. (\ref{Rt}) goes over on one sublattice only.  
We also followed the time-dependence of the variance of instantaneous frequencies 
\be
\Omega(t)=\frac{1}{L^2}\sum_{i=1}^{L^2}[\dot{\varphi}_i(t)]^2.
\ee
We note that the average of instantaneous frequencies is missing here since, as already mentioned, it is zero due to the absence of intrinsic frequencies. 
As it can easily be shown by using Eq. (\ref{grad}), this quantity is closely related to the energy density $u(t)=U(t)/L^2$ as
\be
\dot u(t)=-\Omega(t).
\label{udot}
\ee
Besides computing the time-dependence of the above quantities, we also considered fixed-point properties of the system. To do so, a practical criterion of convergence was defined as the condition that the largest in magnitude instantaneous frequency is smaller than a predefined limit ($\epsilon=10^{-6}$ unless specified otherwise)  i.e. $\max_i\{|\dot\varphi_i(t)|\}<\epsilon$. The earliest time that this condition is fulfilled in a finite system is identified as the convergence time and denoted by $\tau$. 

The linear system sizes considered in the numerical analysis were integer powers of $2$, typically up to $L=512$, but in certain cases up to $L=4096$. The computations have been carried out for an ensemble of independent random samples (i.e. set of couplings), typically $10^3$ in number for $L\le 256$ and $10^2$ otherwise, with one random initial condition per sample, and finally, an averaging of the above defined quantities was performed. We mention, however, that, according to our checks, taking a single (large) sample and performing computations on it for a set of random initial states does not give significantly different averages from those obtained with an ensemble of random samples.

\section{Numerical results}
\label{sec:results}

\begin{figure}[h]
\includegraphics[width=80mm]{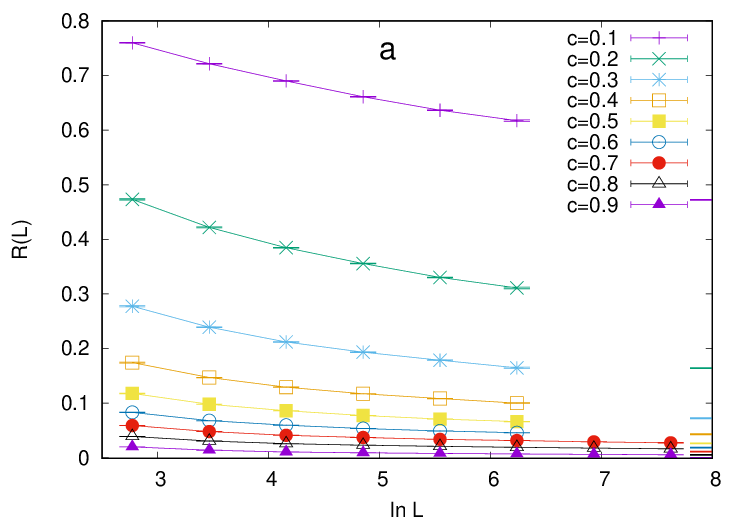}
\includegraphics[width=80mm]{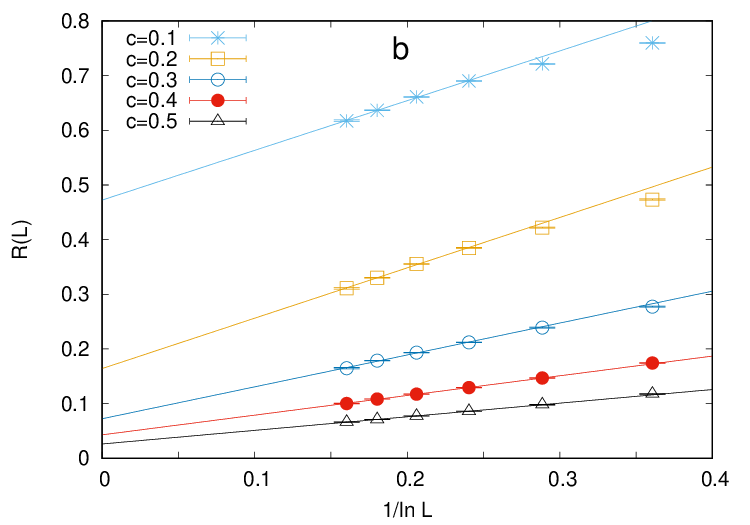}
\includegraphics[width=80mm]{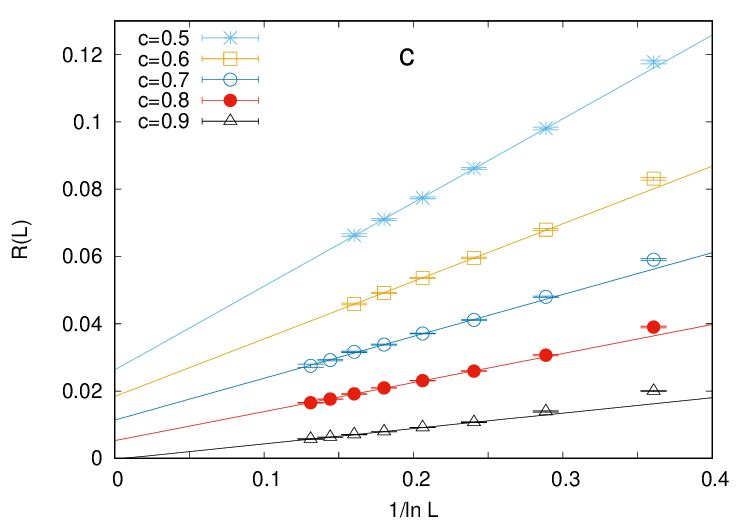}
\caption{\label{fssfig} 
Size dependence of the average fixed-point ($t\to\infty$) order parameter 
obtained numerically for the binary distribution of couplings with different $c$ (a). The same data are plotted against $1/\ln L$ in panel b and c, separately, for a better visibility. The straight lines are linear fits to the data in the domain $1/\ln L<0.25$.      
}
\end{figure}

\begin{figure}[h]
\includegraphics[width=80mm]{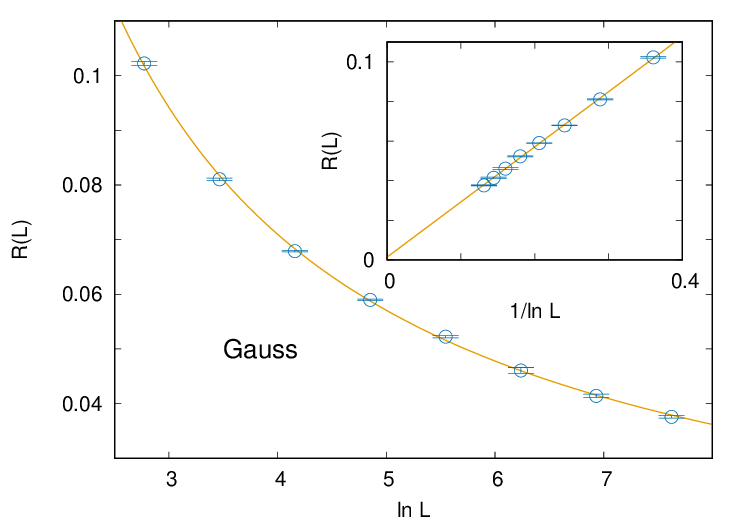}
\caption{\label{fssgfig} 
Size dependence of the average fixed-point ($t\to\infty$) order parameter 
obtained numerically for normally distributed couplings. The same data are plotted against $1/\ln L$ in the inset. The solid line is a linear fit to $R(L)$ vs. $1/\ln L$ in the domain $1/\ln L<0.25$. }
\end{figure}

\begin{figure}[h]
\includegraphics[width=80mm]{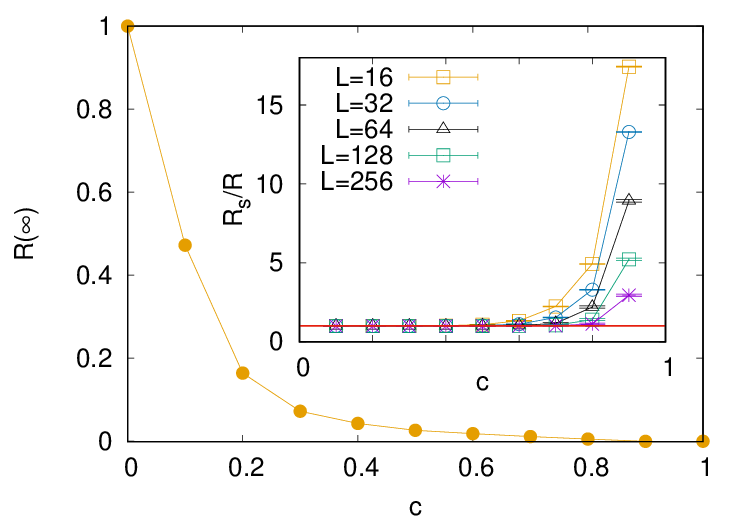}
\caption{
\label{extfig}
The order parameter extrapolated to $L\to\infty$ using the assumption in Eq. (\ref{reclog}), for different values of the parameter $c$ of the binary distribution. 
The inset shows the ratio of the average sublattice order parameter to the total one, for different systems sizes and different values of $c$.
The straight horizontal line is at $R_s/R=1$.
}
\end{figure}

First, we consider the fixed-point properties of the system and their dependence on the system size. We found that, concerning the order parameter and the energy density, the convergence threshold $\epsilon=10^{-6}$ is sufficiently small so that the deviations from the fixed-point values are well within the statistical errors. The average order parameters $R(L)$ obtained  with the binary distribution for a series of concentrations $c$ are plotted against $\ln L$ in Fig. \ref{fssfig}a, while the data for the Gaussian distribution are shown in Fig. \ref{fssgfig}. 
As can be seen, the order parameter converges in all cases very slowly with the system size, and, as it is demonstrated in Fig \ref{fssfig}b, \ref{fssfig}c, and the inset of Fig. \ref{fssgfig}, the data fit well to a reciprocal logarithmic law of the form
\be
R(L)-R(\infty)\simeq\frac{a}{\ln L}
\label{reclog}
\ee
for large $L$, with some non-universal constants $a$ and $R(\infty)$. 
The latter constant, the order parameter in the limit $L\to\infty$, as obtained by a linear fit in terms of $1/\ln L$, is close to zero [$0.001(1)$] in the case of the Gaussian distribution of couplings, see also the inset of Fig \ref{fssgfig}. For the binary distribution, however, as it is shown in Table \ref{table} and Fig \ref{extfig}, the limiting values $R(\infty)$ obtained under the assumption of Eq. (\ref{reclog}) are significantly greater than zero for not too high values of the concentration of negative couplings. 

Next, we address the question whether the sublattice order parameter $R_s$ is significantly different from the total order parameter $R$. This may be expected at a high concentration of negative bonds, since in the case $c=1$, there is an antiferromagnetic ordering at the stable fixed point with $R=0$ and $R_s=1$.
As can be seen in the inset of Fig \ref{extfig}, the ratios $R_s(L)/R(L)$ of finite-size data are indeed greater than one for high values of $c$. Nevertheless, with increasing system size, they steadily decrease, 
the upper bound of the domain in which $R_s/R\approx 1$ gradually shifting to higher and higher values of $c$. This suggests that for $c<1$, the antiferromagnetic ordering presumably does not survive the limit $L\to\infty$.

\begin{figure}[h]
\includegraphics[width=80mm]{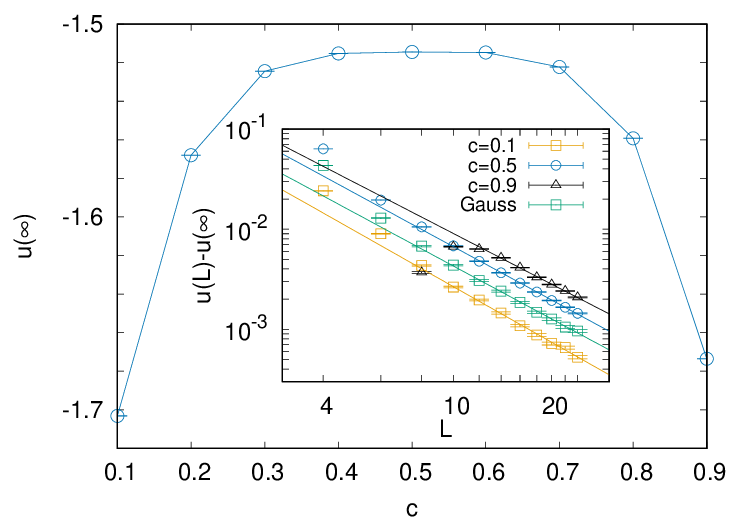}
\caption{
\label{enfig}
Main figure. The average energy density $u=U/L^2$ at the fixed point for the binary distribution with different values of $c$. The data, obtained for system size $L=256$, give essentially the infinite-size limit of $u$.
Inset. Finite-size scaling of the energy density for different coupling distributions. The straight lines are linear fits to the data; the estimated exponents can be found in Table \ref{table}. 
}
\end{figure}

We also computed the average energy density $u=U/L^2$ of the corresponding XY model at the stable fixed point, see Eq. (\ref{UXY}). Due to the rapid convergence of $u$ with the system size, the data obtained for $L=256$ are very close to their $L\to\infty$ limiting values. For the Gaussian distribution, we obtained $-1.3524(1)$, while for the binary distribution, the data can be found in Table \ref{table} and are plotted in Fig \ref{enfig} for different values of $c$.  The energy density takes its maximal value $-1.5145(1)$ at $c=0.5$, which is somewhat higher than the ground-state energy $-1.55331(56)$ of the same model obtained in Ref. \cite{weigel}. 
As it is demonstrated in the inset of Fig \ref{enfig}, the convergence of finite-size energy densities is algebraic, 
\be
u(L)-u(\infty)\sim L^{-\kappa}.
\label{uL}
\ee 
Estimates of the exponent $\kappa$ obtained by power-law fits to the data for some selected coupling distributions can be found in Table \ref{table}. 
We note that, at a $T=0$ critical point, the ground-state energy density should scale as in Eq. (\ref{uL}), with an exponent $\kappa=d-\Theta_s$, where $d$ is the spatial dimension and $\Theta_s$ is the so-called spin stiffness exponent of droplet theory, see Ref. \cite{weigel} and references therein. 
There, the estimate $\Theta_s=-0.329(14)$ was given for $c=0.5$. In the case of the metastable states studied in this paper, we find significantly different exponents $\kappa$, so that $\Theta_s=d-\kappa$ is positive.    
 
\begin{figure}[h]
\includegraphics[width=80mm]{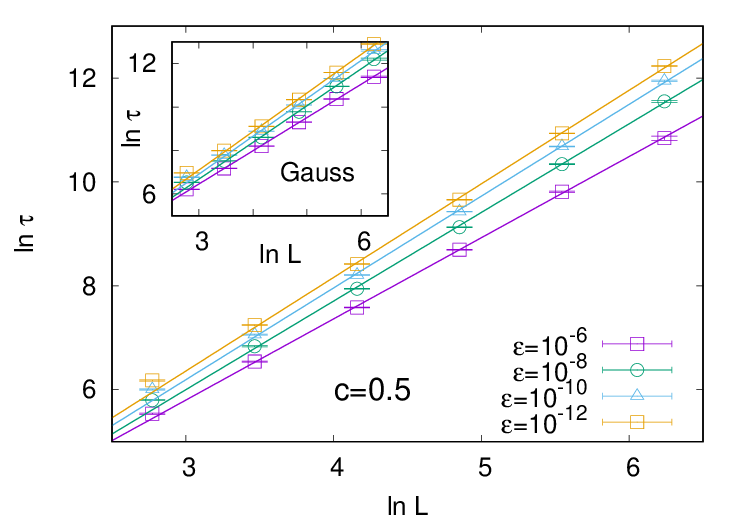}
\caption{\label{taufig} Finite-size scaling of the average relaxation time $\tau$ for the binary distribution with $c=0.5$ (main figure), as well as for the Gaussian distribution (inset) for different values of convergence thresholds $\epsilon$. The straight lines are linear fits to the data in the domain $L\ge 32$.  
}
\end{figure}

Now, we turn to the dynamics of the model. The average convergence (or relaxation) time as a function of the system size $L$ is shown for the binary distribution with $c=0.5$ and for the Gaussian distribution in a double logarithmic plot in Fig \ref{taufig}. 
As can be seen in the figure, the data fit well to a power law
\be 
\tau(L) \sim L^z
\label{pl}
\ee
in both cases. But, to obtain a correct estimate of the dynamical exponent $z$, it is insufficient to use $\epsilon=10^{-6}$, as the slopes show a considerable variation with decreasing $\epsilon$ even beyond this value. The extrapolation of the slopes to $\epsilon=0$ yields $z=1.88(8)$ for the binary distribution and $z=1.84(7)$ for the Gaussian distribution.   
\begin{table}[h]
\begin{center}
\begin{tabular}{|c|c|c|c|}
\hline
 disorder type    &   $R(\infty)$   & $u(\infty)$   &$\kappa$ \\
\hline
\hline
Gaussian    & 0.001(1) &  -1.3524(1) & 1.76(5)   \\ 
\hline
$c=0.1$  &  0.472(4)  &   -1.7032(1) & 1.85(8)  \\     
\hline
$c=0.2$  &  0.164(5) &    -1.5680(1) & \\      
\hline
$c=0.3$  &  0.072(4) &    -1.5244(1) & \\        
\hline
$c=0.4$  &  0.043(2) &    -1.5153(1) &\\      
\hline
$c=0.5$  &  0.0263(6) &   -1.5145(1) & 1.77(4)  \\      
\hline
$c=0.6$  &  0.0184(2) &   -1.5148(1) & \\      
\hline
$c=0.7$  &  0.0114(4) &   -1.5223(1) & \\     
\hline
$c=0.8$  &  0.0052(2) &   -1.5592(1) & \\       
\hline
$c=0.9$  & -0.0003(1) &   -1.6736(1) & 1.69(4) \\     
\hline
\end{tabular}
\end{center}
\caption{\label{table} 
The extrapolated values $R(\infty)$ of the order parameter obtained by the conjectured finite-size dependence in Eq. (\ref{reclog}), the estimated limiting energy densities $u(\infty)$, and the estimated exponents $\kappa$ in Eq. (\ref{uL}) for the Gaussian distribution and the binary distribution with various values of $c$.}
\end{table}

\begin{figure}[h]
\includegraphics[width=80mm]{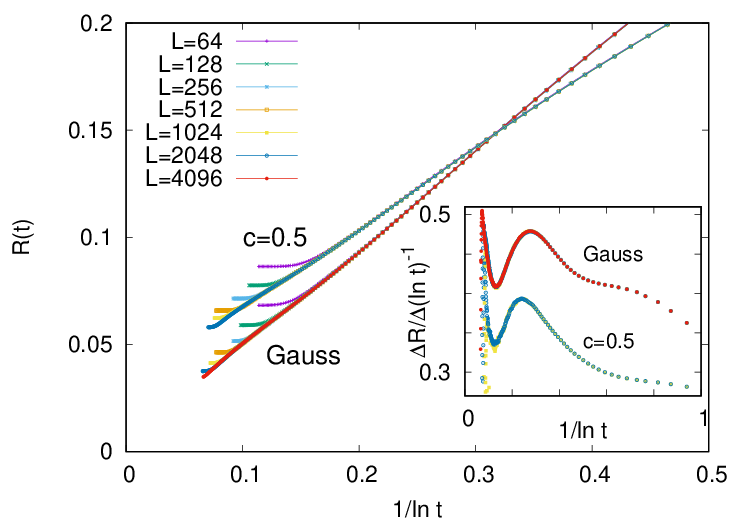}
\caption{\label{Rtfig} 
The time-dependence of the average order parameter for the Gauss distribution and for the binary distribution of couplings with $c=0.5$, for different system sizes. The inset shows the local slopes of the curves defined in the text.  
}
\end{figure}

The conjectured form of the finite-size dependence of the order parameter in Eq. (\ref{reclog}), together with the power-law scaling of the relaxation time with the system size in Eq. (\ref{pl}) imply that, in an infinite system, the average Kuramoto order parameter tends for late times to its limiting value according to a reciprocal logarithmic law as
\be
R(t)-R(\infty)\simeq\frac{az}{\ln t}.
\label{Rt}
\ee
As it is shown in Fig \ref{Rtfig} in two representative cases of coupling distributions, the time dependence of the order parameter indeed follows the law in Eq. (\ref{Rt}) for long times. As can be observed already in the main figure, on top of the reciprocal logarithmic trend, a weak and slow temporal modulation also appears. This is better seen in the inset of the figure, which displays the temporal variation of the discrete local slope of the curves, $\frac{\Delta R(t)}{\Delta (\ln t)^{-1}}$ calculated from adjacent data points. This means that the constant $a$ in Eq. (\ref{Rt}), as well as in Eq. (\ref{reclog}) should be replaced by a slowly varying function $a(t)$. We note that the modulation, although less visibly, appears also in the finite-size dependence of the fixed-point order parameter. 
The late-time fate of this modulation i.e. whether it develops to steady oscillations or dies away can hardly be predicted from the limited-time data at our disposal. 

\begin{figure}[h]
\includegraphics[width=80mm]{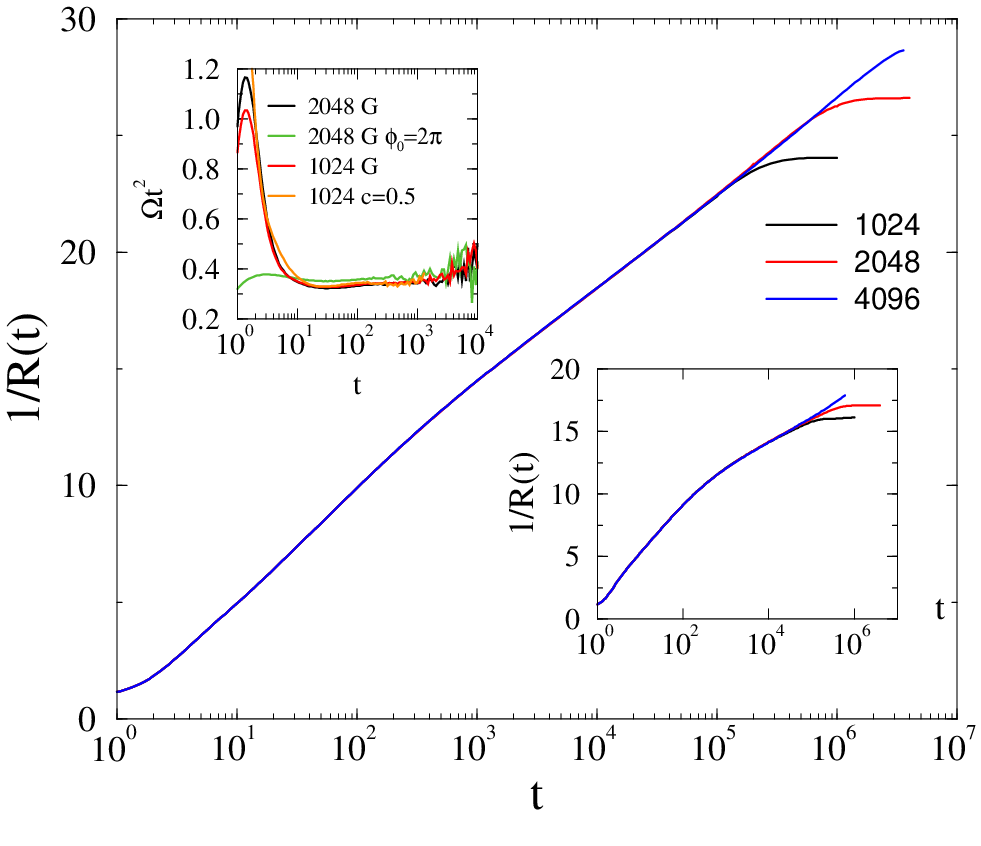}
\caption{\label{Rinvfig} Time-dependence or the average order parameter for different sizes for the Gaussian distribution (main figure) and the binary distribution with $c=0.5$ (lower right inset). The upper left inset shows the rescaled variance of instantaneous frequencies for Gaussian distribution (G), also with an unsynchronized initial condition ($\phi_0=2\pi$) and for the binary distribution with $c=0.5$.}
\end{figure}

An alternative plot of the time-dependence, $1/R(t)$ against $\ln t$, which was also used in Ref. \cite{daido} in the case of all-to-all coupling is shown in Fig. \ref{Rinvfig}. 
Plotting the time-dependence of $R(t)$ in this way gives a linear graph as far as $R(t)\gg R(\infty)$, which is fairly well fulfilled for the Gaussian case where $R(\infty)$ is close to zero but gives limitations in the case of binary distribution, see Table \ref{table}. 

Finally, the time-dependence of the variance $\Omega(t)$ of instantaneous frequencies for two representative cases, and, in the case of Gaussian couplings, also for an unsynchronized initial state with $\phi_0=2\pi$, is shown in the upper left inset of Fig \ref{Rinvfig}.
The numerical data in all cases are compatible with an algebraic temporal decay 
\be 
\Omega(t)\sim t^{-\alpha}
\label{omega_t}
\ee
with an exponent $\alpha$ close to $2$ (the estimated values are   
$\alpha=2.01(3)$ for the binary distribution with $c=0.5$ and $\alpha=1.98(3)$ for the Gaussian distribution). This is demonstrated indirectly in the figure by plotting $\Omega(t)t^2$ against $t$, displaying a roughly constant value at late times. 
Eq. (\ref{udot}) then implies that the time-dependent energy density must tend to its limiting value as 
\be
u(t)-u(\infty)\sim t^{-(\alpha-1)}
\ee
at late times. Comparing this with Eqs. (\ref{uL}) and (\ref{pl}), we obtain the relation 
\be 
\alpha-1=\frac{\kappa}{z},
\ee
which is compatible with the estimates of different exponents. 

\section{Discussion}
\label{sec:discussion}

In this paper, we studied the fixed-point and relaxation properties of a two dimensional lattice of identical oscillators with frustrated couplings.  According to our numerical results, the order parameter at the stable fixed point which is reached starting from sightly perturbed synchronized states shows generally a slow, $O(1/\ln L)$ convergence with the system size. The limiting value as $L\to\infty$ is found to depend on the type of coupling distribution. For Gaussian couplings, it is close to zero, whereas for binary distributions, for a sufficiently high concentration of positive (or ferromagnetic in the language of the XY model) couplings, it is significantly above zero. 
This feature of the fixed point which is a metastable state of the XY model is different from that of the the true ground state, where the numerical results quoted in the Introduction point rather toward a vanishing magnetization for $c>0$ in the $L\to\infty$ limit. 
The relaxation time is found to show a power-law dependence on the size, with dynamical exponents having at most a weak dependence on the coupling distribution.  
These findings imply, in consistence with our direct numerical results, that the order parameter in an infinite system approaches its limiting value proportionally to $1/\ln t$ at late times. We have thus demonstrated that the superslow relaxation observed in the model with all-to-all coupling in Ref. \cite{daido}, appears also in dimensions as low as $d=2$. On the top of this logarithmic trend, we also observed a weak, slowly varying modulation in the time-dependence of the order parameter but its persistence at long times cannot be decided from the present numerical data. 
The variance of the instantaneous frequencies is found to decrease algebraically with an exponent close to $-2$. Assuming that the exponent is strictly $-2$, this means that the typical instantaneous frequencies at time $t$ are $O(t^{-1})$, thus the typical variations of the phase variables up to time $t$ are at most $O(\ln t)$. It is tempting to say that this slow variation of the phases alone explains the observed logarithmically slow relaxation of the order parameter. But it is not the case since, as we have seen, an other quantity, the energy density approaches its limiting value algebraically in time.  

The slow temporal decay of the order parameter observed in this work in a model system is reminiscent of the slow decrease of the remanent magnetization measured in quench experiments of spin glass materials \cite{binder}. Here, the system is cooled down below the spin glass temperature in the presence of a strong magnetic field, and after switching off the field, the measurement of the magnetization starts. From the theoretical side, the droplet theory predicts the remanent magnetization to decay proportionally to a negative power of $\ln t$ \cite{fisher_huse}, while Monte Carlo simulations of Ising spin glasses showed a power-law decay with temperature-dependent exponents at non-zero temperatures \cite{rieger,kisker}. We note, however, that the phenomenological description of such a process at zero temperature is not obvious.  

Finally, one may pose the question, what the fate of the observed slow relaxation of the order parameter is in the more general Kuramoto model in Eq. (\ref{gen}) i.e. in the presence of quenched random intrinsic frequencies $\omega_i$.
This has been examined in the frustrated model with all-to-all couplings in a series of works, with the conclusion that slow relaxation gives way in this system to a much more rapid, algebraic or exponential temporal decay, depending on the strength of couplings \cite{daido_prl,stiller,daido_pre}.   
\begin{figure}[h]
\includegraphics[width=80mm]{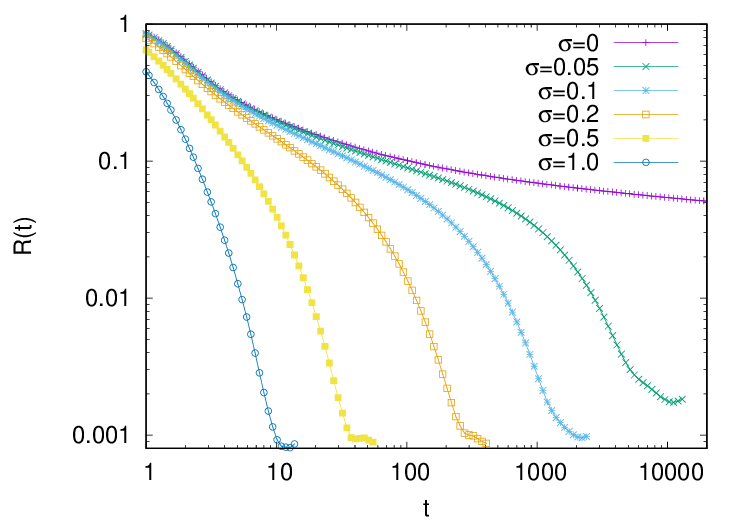}
\caption{\label{rofig} Time-dependence or the average order parameter in the general model in Eq. (\ref{gen}) with Gaussian coupling distribution. The intrinsic frequencies $\omega_i$ are drawn from a zero-mean normal distribution with different standard deviations $\sigma$, the case $\sigma=0$ corresponding to the model studied throughout this work. The size of the system was $L=1024$.}
\end{figure}
As our preliminary results on the model in Eq. (\ref{gen}) show in Fig \ref{rofig}, the slow relaxation of the order parameter is destroyed by the presence of intrinsic frequencies also in two dimensions, the temporal decay becoming faster than algebraic. The detailed study of this model is left for future research.

\begin{acknowledgments}
We thank José A. Hoyos for useful discussions. 
This work was supported by the National Research, Development and Innovation Office NKFIH under Grant No. K146736. We thank 
KIFU for the usage of the Hungarian supercomputer Komondor.
\end{acknowledgments}

\section*{Data Availability Statement}

The data that support the findings of this study are available from the corresponding author upon reasonable request.

\appendix

\bibliography{bib}

\end{document}